\documentclass[aps,prl,twocolumn,showpacs,floatfix,superscriptaddress]{revtex4}
\usepackage{graphicx,bm,amsmath,amssymb,natbib,url,epsfig,psfrag}
\usepackage{color}

\begin{document}

\title{Absence of Aharonov-Bohm Effect of Chiral Majorana Fermion Edge States}
\author{Sunghun Park}
\affiliation{Department of Physics, Korea Advanced Institute of Science and Technology, Daejeon 305-701, Korea}
\author{Joel E. Moore}
\affiliation{Department of Physics, University of California, Berkeley, California 94720, USA}
\affiliation{Materials Sciences Division, Lawrence Berkeley National Laboratory, Berkeley, California 94720, USA}
\author{H.-S. Sim}  \email{hssim@kaist.ac.kr}
\affiliation{Department of Physics, Korea Advanced Institute of Science and Technology, Daejeon 305-701, Korea}

\date{\today}

\begin{abstract}
Majorana fermions in a superconductor hybrid system are charge neutral zero-energy states. For the detection of this unique feature, we propose an interferometry of a chiral Majorana edge channel, formed along the interface between a superconductor and a topological insulator under an external magnetic field. The superconductor is of a ring shape and has a Josephson junction that allows the Majorana state to enclose continuously tunable magnetic flux. Zero-bias differential electron conductance between the Majorana state and a normal lead is found to be independent of the flux at zero temperature, manifesting the Majorana feature of a charge neutral zero-energy state. In contrast, the same setup on graphene has no Majorana state and shows Aharonov-Bohm effects.
\end{abstract}
\pacs{71.10.Pm, 73.23.-b, 74.45.+c, 74.90.+n}


\maketitle


{\it Introduction.---}
There have been efforts to find the evidence of Majorana fermions in superconductor hybrid systems~\cite{Kitaev,Hasan,Alicea,Flensberg}.
In the systems, Majorana fermions appear at zero excitation energy in superconducting energy gap. 
Recent experiments~\cite{Mourik,Das,Deng,Rokhinson} studied a superconductor coupled to a semiconductor nanowire with strong spin-orbit coupling. The result such as zero-bias resonant tunneling agrees with the behavior of a Majorana bound state~\cite{Oreg,Lutchyn} formed at the end of a topological superconducting region. Other experiments~\cite{Williams} found anomalous Fraunhofer diffraction pattern in a Josephson junction on a topological insulator (TI)~\cite{Sacepe,Veldhorst}. This may be related to a Majorana state~\cite{Fu}, however, more studies are necessary to understand it. There are also other proposals~\cite{Fu_interf,Akhmerov_interf,Law,Diez,Wieder,Beri,Tanaka,Li}, including $Z_2$ interferometers~\cite{Fu_interf,Akhmerov_interf,Law} formed along a superconductor-ferromagnet interface on a TI. 


To achieve more direct evidence, it needs to explore other Majorana features.
One goal of the present work is to develop an experimentally feasible test for a Majorana fermion as a {\em charge neutral} zero-energy state, based on an Aharonov-Bohm interferometer; a similar strategy was adapted~\cite{Greenberger} to experimentally confirm the charge neutrality of neutrons. As a charge neutral particle, Majorana fermions will not show Aharonov-Bohm effects. 
This 
is a direct consequence of the fact that Majorana fermions ``are their own antiparticles'', namely, that the Majorana operator is self-conjugate or real, not carrying a complex Aharonov-Bohm phase factor.




It is nontrivial to find an interferometry where Majorana states do not show Aharonov-Bohm effects, because of a few reasons. First, the interferometry has to be formed solely by extended Majorana channels~\cite{Fu_interf,Akhmerov_interf,Law}. When an interference loop enclosing magnetic flux is composed of Majorana bound states and electron paths, Aharonov-Bohm effects occur~\cite{Fu_tele}. It is because a Majorana state in solids is a superposition of an electron and a hole, hence, the tunneling between a Majorana state and an electron path carries flux information. Second, the interferometry should enclose continuously tunable magnetic flux. Its candidate is a superconducting ring with a Josephson junction, rather than a closed ring~\cite{Beri} enclosing quantized flux. 
This setup was recently studied~\cite{Diez,Wieder} in a different context from our study.



\begin{figure}[b]
\includegraphics[width=0.42\textwidth]{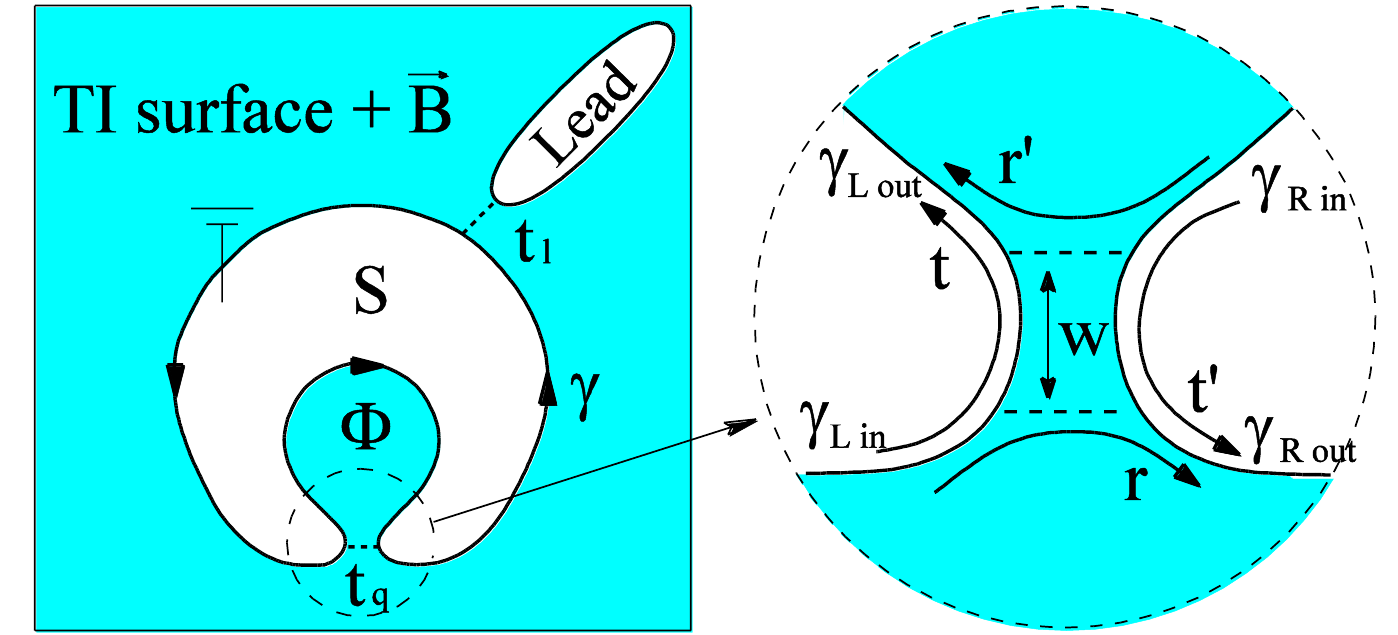}
\caption{Interferometer of a chiral Majorana fermion edge channel $\gamma$ (arrows), formed along the interface between a s-wave superconductor ring (S) and the integer quantum Hall state of a topological-insulator (TI) surface under a magnetic field $\vec{B}$. It has a Josephson junction (dashed circle, right panel), which behaves as a beam splitter for the Majorana channel and allows the channel to enclose continuously tunable magnetic flux $\Phi$. Zero-bias electron tunneling differential conductance between a normal lead and the ring shows that the Majorana state is independent of $\Phi$.}
\label{fig:IF}
\end{figure}

In this work, we propose a Majorana version of quantum Hall interferometers~\cite{Ji} for detecting a Majorana state; see Fig.~\ref{fig:IF}. It is based on a chiral Majorana fermion edge channel, formed along the interface between a s-wave superconductor ring and the integer quantum Hall state of a TI surface. The ring has a Josephson junction, which connects the Majorana channels of the inner and outer boundaries of the ring. It allows the Majorana state to enclose continuously tunable magnetic flux $\Phi$. Zero-bias electron tunneling differential conductance from a normal lead to the ring is found to be independent of $\Phi$ at zero temperature, demonstrating the unique Majorana feature of a charge neutral zero-energy state. 
The setup also exhibits Majorana features at finite bias and temperature. To prove that this behavior is a sensitive probe of Majorana physics, we show that in the same setup on graphene, which also has zero-energy states (because of Berry phase $\pi$ of its Dirac fermions) but has Bogoliubov (a complex superposition of an electron and a hole, such as quasiparticles in a typical superconductor) rather than Majorana fermions, Aharonov-Bohm effects are present.
We discuss the experimental feasibility of our setup.



{\it Chiral Majorana edge channel.---} 
In Fig.~\ref{fig:IF}, a magnetic field $B$ is perpendicularly applied to the TI surface outside the proximity region with superconducting gap $\Delta_0$. The resulting Landau-level orbits of the surface undergo Andreev reflections~\cite{Akhmerov_graphene} and form chiral Majorana edge channels along the interface between the superconducting-gap region and the Landau-gap region~\cite{Tiwari,Park}; a Majorana channel can be also formed when a Zeeman gap induced by a ferromagnet~\cite{Fu_interf,Akhmerov_interf,Law,Tanaka} replaces the Landau gap. The number of the channels is odd (since the TI has an odd number of Dirac cones), indicating that the Majorana states are stable, and it is determined by $B$, the chemical potential $\mu$~\cite{Tiwari}, and $g$-factor $g$~\cite{Park} of the TI surface. Hereafter we focus on the case of a single Majorana channel $\gamma$ well localized near the interface, and on energy scales $\ll \Delta_0$. The particle-hole symmetry ensures that $\gamma$ satisfies $\gamma(x) = \gamma^\dagger (x)$ and $\gamma_k^\dagger = \gamma_{-k} \sim \int dx \gamma (x) e^{-ikx}$, with coordinate $x$ and momentum $k$ along the channel. Its Hamiltonian is 
\begin{eqnarray}
H_{MF} = -i \hbar v_M \int d x \gamma (x) \partial_{x} \gamma (x).
\end{eqnarray}
$v_M = v_M (B, \Delta_0, \mu, g)$ is the propagation velocity of $\gamma$.

The Josephson junction in Fig.~\ref{fig:IF} is in the short junction limit. It describes the coupling of Majorana channels $\gamma_{L,R}$ between its two sides. Its model Hamiltonian~\cite{Fu} is
\begin{equation}
H_{JJ} = -2 i \int d x t_q (x) \gamma_L(x) \gamma_R(x). \label{HamiltonianJJ}
\end{equation}
$t_q (x) = \Delta_0 \cos (\phi /2)$ for $| x - x_0 | \le W/2$, and $t_q (x) = 0$ otherwise, where $x_0$ and $W$ are junction center and width. Superconducting phase difference $\phi$ across the junction is induced by $\Phi$ as $\phi/2 = 2 \pi \Phi /\Phi_{0,e}$ with $\Phi_{0,e} = h/e$. The junction behaves as a beam splitter of $\gamma_k$. The channel $\gamma_{k, \textrm{in}}$ with momentum $k$, incoming to the junction from the outer or inner boundary of the ring, is scattered into outgoing ones $\gamma_{k, \textrm{out}}$. We obtain the unitary scattering matrix $S_{JJ}$ for this [see Fig.~\ref{fig:IF}],
\begin{eqnarray}\label{SJJ}
\left( \begin{array}{c} \gamma_{L, \textrm{out}} \\  \gamma_{R, \textrm{out}} \end{array} \right) = S_{JJ}
\left( \begin{array}{c} \gamma_{L, \textrm{in}} \\  \gamma_{R, \textrm{in}} \end{array} \right),
\,\,\,\,\,
S_{JJ}= \left( \begin{array}{cc} t & r' \\ r & t' \end{array} \right), \label{SMATRIX}
\end{eqnarray}
reflection coefficient $r = -r' = (\eta^{-1} \sinh \alpha) \Delta_0 \cos \frac{\phi}{2} $, and transmission coefficient
$t = t' = \hbar v_M \alpha / (\eta W)$, where $\alpha = W \sqrt{[\Delta_0 /(\hbar v_M)]^2 \cos^2(\phi/2)- k^2}$ and $\eta = \hbar v_M (\alpha W^{-1} \cosh \alpha - i k \sinh \alpha)$; the same expression was found in Ref.~\cite{Wieder}.
Note that $r$ and $t$ depend on $\phi$, hence, on $\Phi$ in a nontrivial way; when $\phi = \pi$ and $k \to 0$, Majorana states occur in the junction so that $r \to 0$ and $t \to 1$.
We will see that the flux dependence does not affect the Majorana resonance state at $k=0$ in our setup.  

 

{\it Resonance.---} We study scattering between the incoming and outgoing channels of the outer ring boundary, $\gamma_{L, \textrm{in}}$ and $\gamma_{R, \textrm{out}}$, at the Josephson junction. Because there is no loss of Majorana fermions along the inner boundary of the ring, the scattering causes phase shift $\theta_\Lambda$ only. From Eq.~\eqref{SMATRIX} and $\gamma_{R, \textrm{in}} = e^{ikd} \gamma_{L, \textrm{out}}$ (the accumulation of dynamical phase $kd$ along the circumference $d$ of the inner ring boundary), we obtain $\gamma_{R, \textrm{out}} = e^{i \theta_{\Lambda}} \gamma_{L, \textrm{in}}$,
\begin{equation}\label{PhaseLambda}
e^{i \theta_\Lambda}= r + \frac{t t' e^{i k d}}{1-r' e^{i k d}} = \frac{r + (r^2 + t^2) e^{i k d}}{1 + r e^{i k d}}
\end{equation}
for $|r| \ne 1$.
The first equality of Eq.~\eqref{PhaseLambda} comes from the direct scattering between $\gamma_{L, \textrm{in}}$ and $\gamma_{R, \textrm{out}}$ and from the paths with multiple winding of the flux $\Phi$ along the inner boundary, while the second from the unitarity of $S_{JJ}$. 

We notice a number of interesting points from Eq.~\eqref{PhaseLambda}. First, $\theta_\Lambda$ depends on $\Phi$ in a nontrivial way through $r(\Phi)$ and $t(\Phi)$. This is distinct from usual electron cases where the flux dependence couples with dynamical phase such as $kd + 2 \pi \Phi / \Phi_{0,e}$. 
Second, at zero energy (i.e., $k=0$), the followings are satisfied, irrespective of $\Phi$: $r_0 (\Phi) \equiv  r(k = 0) = \tanh (\frac{W \Delta_0}{\hbar v_M} \cos \frac{2 \pi \Phi}{\Phi_{0,e}})$ is real, $r^2 + t^2 = 1$ (partially from the unitarity of $S_{JJ}$), thus, $e^{i \theta_\Lambda} = 1$ is real. These are attributed to the fact that the Majorana operator is real self-conjugate.
Third, near zero energy ($k \simeq 0$), $r^2 + t^2 \simeq 1$ and $r \simeq  |r_0(\Phi)| \textrm{sgn}(\cos 2 \pi \Phi / \Phi_{0,e})$ are almost real~\cite{Deviation_real}. Then, when $kd = m \pi$ with $m=1,2,\cdots$ is satisfied, the phase shift becomes $\theta_\Lambda \simeq m \pi$, almost independent of $\Phi$. Considering the resonance condition $kd + \pi + \textrm{arg} \, r = 2 m' \pi$ (with integer $m'$) of the inner boundary and $\textrm{arg} \, r = \pi [1- \textrm{sgn} (\cos 2 \pi \Phi / \Phi_{0,e})]/2$, we find that $kd = m \pi$ means on or off resonance of the inner boundary, depending on $\Phi$. However, regardless of on or off resonance, $\theta_\Lambda$ is almost independent of $\Phi$. In contrast, for $kd \ne m \pi$, $\theta_\Lambda$ varies with $\Phi$.
Fourth, Eq.~\eqref{PhaseLambda} describes well the vortex limit where the ring is fully closed. In this case, $\Phi$ will be quantized as $\Phi = l \Phi_{0,e} / 2$ with integer $l$ so that $r = (-1)^l$ and $t=0$, hence, $e^{i\theta_\Lambda} = (-1)^l$. 

The resonance condition of the whole setup (including both the inner and outer boundaries) is found as
\begin{eqnarray}
k L + \pi + n_v \pi + \theta_{\Lambda} = 2 n \pi, \label{resonance_condition} \end{eqnarray}
where $kL$ is the dynamical phase along the circumference $L$ of the outer boundary, $\pi$ is the Berry phase~\cite{Fu_interf,Akhmerov_interf,Law} of Majorana fermions circulating the setup, $n_v$ is the number of vortices inside the superconducting area, and $n$ is an integer.


\begin{figure}
\includegraphics[width=0.47\textwidth]{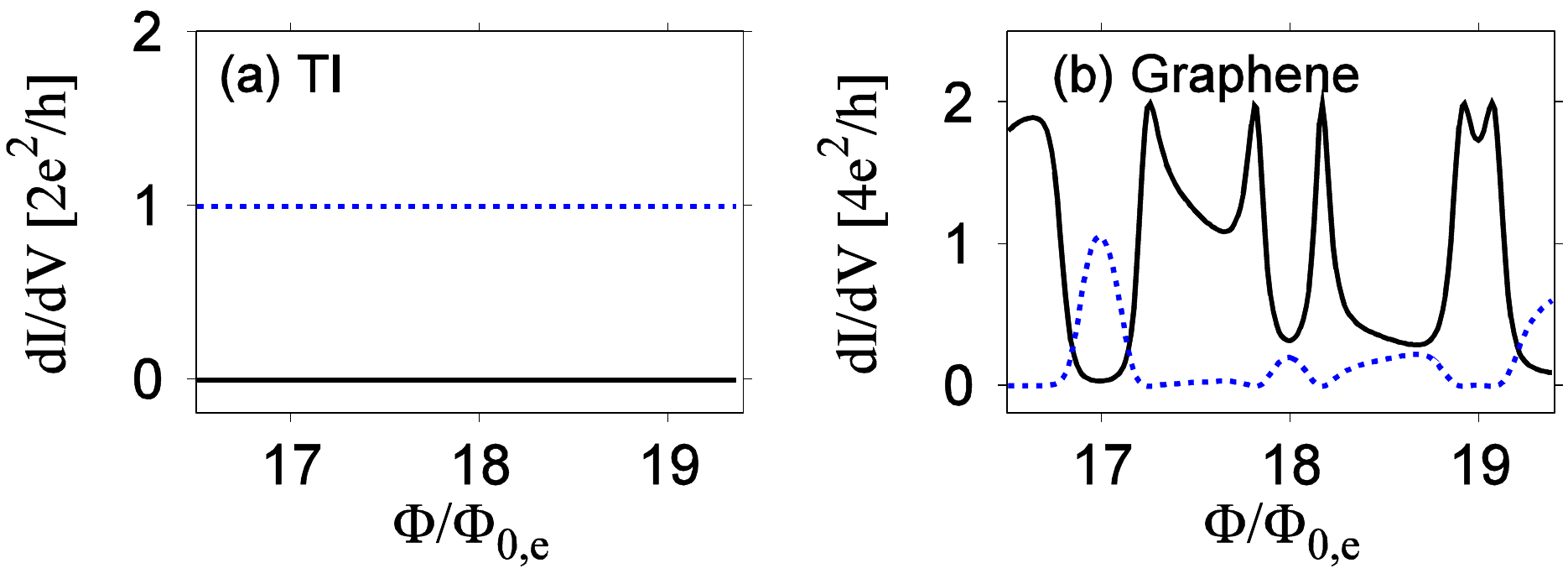}
\caption{(a) $dI/dV$ of the setup on a TI surface (see Fig.~\ref{fig:IF}) at zero temperature and at zero bias, as a function of the magnetic flux $\Phi$, for the cases of odd $n_v$ (blue dashed curve) and even $n_v$ (black solid).
(b) The same as in (a), but for the setup on graphene. Both in (a) and (b), $B$ changes from 0.85 T to 1 T, and we choose $\Delta_0=1.5$ meV, $L=7000$ nm, $d=1000$ nm, $g$-factor $g=2$, chemical potential $\mu=15$ meV, and Fermi velocity $v_f = 5 \times 10^5$ m/s; these parameters lead to $v_M = 0.14 v_f$. We also choose $W$ such that the maximum value of $|r|^2$ is 0.6 for (a) and 0.8 for (b), and $\tilde{t}_l = $ 0.48 for (a) and $\tilde{t}_l =$ 0.62 for (b). The Majorana state in case (a) does not exhibit Aharonov-Bohm effects, while the zero-energy Bogoliubov excitations in graphene case (b) do.
}
\label{fig:dIdVzeroE}
\end{figure}

The resonances can be detected by electron tunneling from a normal lead to the outer ring boundary. The lead is modeled by one-dimensional electrons with Hamiltonian $H_L = -i \hbar v_L \sum_{\sigma = \uparrow, \downarrow } \int^{\infty}_{-\infty} d y \psi^{\dagger}_{\sigma}(y) \partial_y \psi_{\sigma}(y)$, where $\psi^{\dagger}_{\sigma}(y)$ is the electron field operator with spin $\sigma$ at position $y$ and $v_L$ is electron velocity in the lead. The tunneling Hamiltonian is $H_{MF-L} = -2 i t_l \gamma (x_1) \bar{\gamma} (y_1)$, where tunneling strength $t_l$ is real, $x_1$ and $y_1$ are tunneling positions,
$\bar{\gamma} (y_1) = (1/2) \sum_{\sigma} [ e^{i \chi_{\sigma}} \psi^{\dagger}_{\sigma}(y_1) + e^{- i \chi_{\sigma}} \psi_{\sigma}(y_1)]$ is a Majorana representation of states in the lead, and $e^{i \chi_\sigma}$ is the phase factor from the tunneling. At zero temperature, the differential conductance $dI/dV$ from the lead with bias $V$ to the grounded ring has the form~\cite{Law} of
\begin{equation}\label{dIdVEqn}
\frac{dI}{dV} = \frac{2e^2}{h}|s_{he}|^2 = \frac{2e^2}{h} \frac{\tilde{t}^4_l \text{cos}^2(\theta_s/2)}{\text{sin}^2(\theta_s/2)+\tilde{t}^4_l \text{cos}^2(\theta_s/2)},
\end{equation}
where $s_{he}$ describes Andreev reflections in the lead, $\tilde{t}_l = t_l/(2 \hbar \sqrt{v_M v_L})$, and
$\theta_s = k L + \pi + n_v \pi + \theta_{\Lambda}$. The resonance condition in Eq.~\eqref{resonance_condition} is written as $\theta_s = 2 n \pi$.

We first discuss $dI/dV$ at zero temperature; see Fig.~\ref{fig:dIdVzeroE}. At zero bias, it is determined by the Majorana state with $k=0$, which shows $e^{i \theta_\Lambda} = 1$. Hence, the resonance condition of Eq.~\eqref{resonance_condition} and $dI/dV (V=0)$ do not depend on $\Phi$. This demonstrates the absence of Aharonov-Bohm effects of the Majorana state, the manifestation of the fact that Majorana fermions are charge neutral. $dI/dV (V=0)$ is also independent of system lengths ($L$, $d$, and $W$) and coupling strengths ($t_q$, $t_l$). It shows the $Z_2$ property~\cite{Fu_interf,Akhmerov_interf,Law} that the interferometry has off (on) resonance and $dI/dV = 0$ ($dI/dV = 2e^2 / h$), when $n_v$ is even (odd).

\begin{figure}
\includegraphics[width=0.48\textwidth]{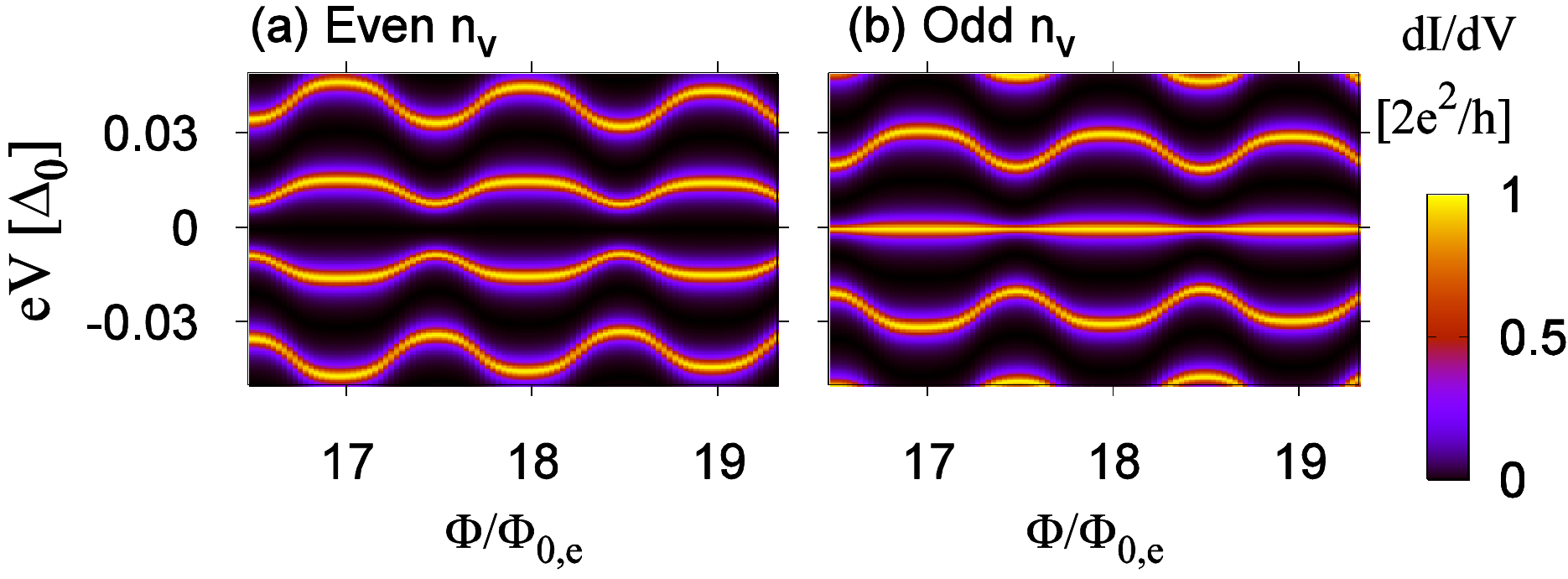}
\caption{
$dI/dV$ of the setup on a TI surface at zero temperature, as a function of $\Phi$ and $V$, for the case of (a) even $n_v$ and (b) odd $n_v$. The same parameters as in Fig.~\ref{fig:dIdVzeroE} are used. 
}
\label{fig:TIAB}
\end{figure}

Next, we discuss $dI/dV$ at zero temperature, but at finite bias; see Fig.~\ref{fig:TIAB}. 
The zero-bias behavior mentioned above appears at $V=0$.
At finite bias, resonances occur whenever Eq.~\eqref{resonance_condition} is satisfied. For low energy of $kd \ll 1$, we find $\theta_\Lambda \simeq \frac{1-r_0}{1+r_0} kd$ and obtain resonance center $V_r$,
\begin{equation}
e V_{r} \simeq  \frac{\hbar v_M (2 n \pi - \pi - n_v \pi)}{L + d(1-r_0)/(1+r_0)}.
\end{equation}
$V_r$ depends on $\Phi$ through $(1-r_0)/ (1+r_0)$, oscillating with period $\Phi_{0,e}$. 
$|V_r|$ has maxima (minima) at (half-)integer multiples of $\Phi_{0,e}$; the gradual decrease of $|V_r|$ with increasing $\Phi$ is due to the dependence of $v_M$ on $B$.
For $kd \ll 1$, the level broadening $\Delta V$ of the resonances also depends on $\Phi$ as 
\begin{equation}
e \Delta V \simeq \frac{2 \hbar v_M \tilde{t}^{2}_l}{L + d (1-r_0)/(1+r_0)},
\end{equation} 
which has maxima (minima) at (half-)integer multiples of $\Phi_{0,e}$. 
This behavior of $V_r$ and $\Delta V$ is the consequence of the Majorana feature of $t_q \propto \cos (\phi / 2)$ in Eq.~\eqref{HamiltonianJJ}. Note that the range of $V$ in Fig.~\ref{fig:TIAB} does not reach the condition of $kd = m \pi$ with nonzero $m$, at which $\theta_\Lambda \simeq m \pi$ and $dI/dV$ is almost independent of $\Phi$ (as discussed above).
We emphasize that at $V=0$ and zero temperature, $dI/dV$ remains constant ($0$ or $2e^2/h$) for finite $\Delta V (\Phi)$ (even if $\Delta V (\Phi)$ is larger than resonance level spacing) for both of even and odd $n_v$.

At finite temperature $T$, one has $dI/dV (V=0) = \frac{2 e^2}{h} \int^{\infty}_{0} d \varepsilon |s_{he}(\varepsilon)|^2 \frac{\beta}{1+\text{cosh} \beta \varepsilon}$~\cite{Fu_interf}, where $\beta = (k_B T)^{-1}$ and $\varepsilon = \hbar v_M k$ is the excitation energy. 
In this case, thermal broadening causes $dI/dV (V=0)$ to depend on $\Phi$. From $|s_{he}|$ in Eq.~\eqref{dIdVEqn}, we find that at $k_B T \ll e \Delta V$ and $\hbar v_M / L$, $dI/dV (V=0)$ weakly depends on $\Phi$ as $dI/dV (V=0) = \frac{2 e^2}{h} [1 - \frac{4 (k_B T)^2}{(e \Delta V)^2} + O(T^4)]$ for odd $n_v$, and as $dI/dV (V=0) \simeq \frac{2 e^2}{h} \frac{4 \tilde{t}^8 (k_B T)^2}{(e \Delta V)^2}$ for even $n_v$.
The dependence on $\Phi$ becomes suppressed as $\sim T^2$ at lower temperature, in sharp contrast to usual electron interferometers~\cite{Ji} where interference visibility becomes enhanced at lower temperature. 
This is a finite-temperature signature of the absence of Aharonov-Bohm effects of the Majorana state. Note that $n_v$ can vary by temperature or $B$ change, leading to jumps of $dI/dV (V=0)$ between 0 and $2e^2/h$~\cite{Rakhmanov,Ioselevich}. The jumps are distinct from Aharonov-Bohm effects as they are not periodic in $\Phi$. 





\begin{figure}
\includegraphics[width=0.47\textwidth]{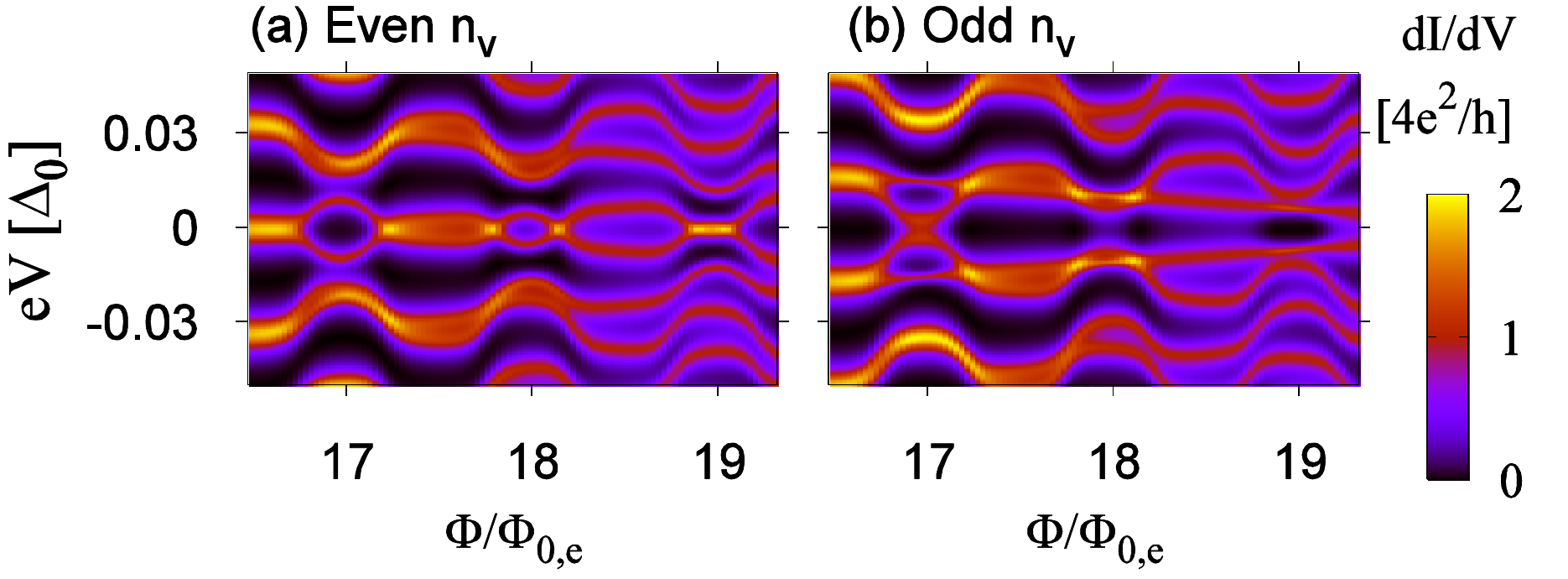}
\caption{
$dI/dV$ of the setup on graphene at zero temperature, as a function of $\Phi$ and $V$, for the case of (a) even $n_v$ and (b) odd $n_v$. The same parameters as in Fig.~\ref{fig:dIdVzeroE} are used. The result is distinct from Fig.~\ref{fig:TIAB}, since there are four chiral channels, each carrying nonzero charge, in the graphene case. 
}
\label{fig:GraAB}
\end{figure}

{\it Non-Majorana case.---} 
To compare the above findings with a non-Majorana case, we consider the same setup on graphene. Similarly to a TI, graphene has Dirac fermions~\cite{Geim}, and zero-energy excitations by superconducting proximity effects. But it has two Dirac cones at K and K' valleys, which are transformed into each other by time reversal. Moreover, the momentum of its Dirac fermions couples to the pseudospin representing the sublattice sites, rather than spin. As a result, near zero excitation energy, there occur four chiral edge channels of $\Psi_1 = (e^{K}_{\uparrow},h^{K'}_{\downarrow})$, $\Psi_2 =(e^{K'}_{\uparrow},h^{K}_{\downarrow})$, $\Psi_3 =(e^{K}_{\downarrow},h^{K'}_{\uparrow})$, and $\Psi_4 =(e^{K'}_{\downarrow},h^{K}_{\uparrow})$ along the interface between the superconducting region and the Landau gap region~\cite{Akhmerov_graphene}, where $\Psi$ represents an electron-hole ($e$-$h$) pair with opposite spin ($\uparrow$, $\downarrow$) and valley. In the same way as the TI case, we compute $dI/dV$, taking into account of the energy dispersion of $\Psi_i$~\cite{Zeeman}. We ignore intervalley mixing and spin scattering in the setup, and neglect minor correction to $S_{JJ}$ by Zeeman energy for simplicity;
these effects do not alter our finding that $dI/dV$ shows Aharonov-Bohm effects at $V=0$ in graphene. 

  
The chiral channels in graphene were theoretically studied in Ref.~\cite{Akhmerov_graphene}, ignoring Zeeman energy. We derive the energy dispersion $\epsilon_i (k) = \hbar v_D k + E_{Z, i}$ of channel $\Psi_{i=1,2,3,4}$, following Ref.~\cite{Akhmerov_graphene}, but including Zeeman energy. $v_D$ is the propagation velocity, and the Zeeman contribution $E_{Z,i}$ is finite and channel dependent~\cite{Park,Zeeman}. 
For zero Zeeman energy, $\Psi_i$ is charge neutral with equal weight between its electron and hole parts, and $dI/dV$ can be independent of $\Phi$ at $V=0$. In contrast, in the realistic case of finite Zeeman energy, $\Psi_i$ carries charge even at zero energy, since its electron and hole parts have opposite spin to each other, hence, have different weight due to Zeeman energy. As a result, $\Psi_i$ does not satisfy the Majorana condition of $\gamma_k^\dagger = \gamma_{-k}$ and $\gamma^\dagger (x) = \gamma (x)$, and $dI/dV (V=0)$ exhibits $\Phi$ dependent oscillation for both of even and odd $n_v$; see Figs.~\ref{fig:dIdVzeroE}(b) and \ref{fig:GraAB}.



{\it Conclusion.---} The absence of Aharonov-Bohm effects at zero bias in our Majorana interferometry is a direct consequence of the essence that a Majorana fermion is its own antiparticle, i.e., a real operator. It is in contrast to the $\Phi$ dependence of $dI/dV$ of the same setup at large bias. It should be also distinguished from Bogoliubov fermions, such as those in the graphene case, that show Aharonov-Bohm effects unless there is fine-tuning (e.g., unrealistic tuning to zero Zeeman energy for graphene).


We discuss experimental feasibility. For the superconducting ring, one may use niobium. 
It has $\Delta_0 \approx 1.5$ meV, superconducting critical temperature of 9 K, the lower and upper critical fields of 2.7 T and 4 T, coherence length $\xi_C \simeq 200$ nm, and penetration depth $\xi_D \simeq 350$ nm~\cite{Rickhaus,Morpurgo2,Oostinga}. It was used to study proximity effects under high magnetic fields in graphene~\cite{Rickhaus} and in a two-dimensional electron gas~\cite{Eroms, Uhlisch}. While $d / (2 \pi)$ should be longer than magnetic length ($\simeq 25$ nm at $B=1$ T), $L$ needs to satisfy $\pi \xi_C, \pi \xi_D < L < \hbar v_M / (k_B T)$, which means $350 \, \textrm{nm} \, < L / \pi < 11500 \, \textrm{nm}$ at $T =$ 12 mK; Majorana resonance energies is resolved in the setup with $L< \hbar v_M / (k_B T)$, we estimate $v_M \simeq 0.14 v_f$~\cite{Tiwari}, and $T \simeq$ 12 mK was achieved~\cite{Williams}. 
This indicates that a range of $L$ is available for our prediction.
Under the parameters ($L = 7000$ nm) used in Fig.~\ref{fig:dIdVzeroE},  
the energy splitting near zero bias is about $0.024 \, \textrm{meV}$ in the Majorana case (see Fig.~\ref{fig:TIAB}), and $0.022 \, \textrm{meV}$ in the graphene case (Fig.~\ref{fig:GraAB}).
Hence, the two cases are distinguishable at currently available temperature of 12 mK ($\simeq$ 0.001 meV).




 
We thank D. Goldharber-Gordon for useful discussions. We acknowledge support by Korea NRF (grant No. 2012S1A2A1A01030312; H.-S. S.) and DARPA (J.E.M).

\section*{Supplemental material : Aharanov-Bohm interferometry of chiral edge states in graphene}

In this material, we derive the low-energy dispersion of chiral edge states 
formed along an interface between a superconductor and the integer quantum Hall state of graphene. 
Next, we provide the form of $dI/dV$ of the interferometry based on graphene, which we used to calculate Fig. 2(b) and 4 in our main text.

\subsection*{1. Hamiltonian of superconductor-graphene junction}

Along an interface between a superconductor and the integer quantum Hall surface of graphene, there are four 
chiral edge channels,     
\begin{widetext}
\begin{equation}
\Psi_1 = \left(
 \begin{array}{c}
   \Psi^{K}_{e \uparrow} \\
   \Psi^{K'}_{h \downarrow}
 \end{array}
\right), \,\,\
\Psi_2 = \left(
 \begin{array}{c}
   \Psi^{K'}_{e \uparrow} \\
   \Psi^{K}_{h \downarrow}
 \end{array}
\right), \,\,\
\Psi_3 = \left(
 \begin{array}{c}
   \Psi^{K}_{e \downarrow} \\
   \Psi^{K'}_{h \uparrow}
 \end{array}
\right), \,\,\
\Psi_4 = \left(
 \begin{array}{c}
   \Psi^{K'}_{e \downarrow} \\
   \Psi^{K}_{h \uparrow}
 \end{array}
\right).
\end{equation}
\end{widetext}
Each channel is formed by a pair of electron and hole with opposite valley ($K, K'$) and spin ($\uparrow, \downarrow$), and is described by 
\begin{equation}\label{G_BdG}
\left(
 \begin{array}{cc}
  H_{\text{G}} - \mu & \,\, \Delta(y) \\
  \Delta^{*}(y) & \,\, \mu - \mathcal{T} H_{\text{G}} \mathcal{T}^{-1} 
 \end{array}
\right)
\Psi_j = 
\epsilon_j
\Psi_j.
\end{equation}
Here, $\mu$ is the chemical potential, $\epsilon_j$ the excitation energy of $\Psi_j$, $\mathcal{T} = i \sigma_y \mathcal{C}$ the time-reversal operator, $\mathcal{C}$ the complex conjugation operator, and $\sigma_{i=x,y,z}$ the Pauli matrices operating on the pseudospin states of Dirac fermions in graphene. $x$ and $y$ are the coordinates along the interface and the axis perpendicular to the interface, respectively. 
Graphene Hamiltonian $H_{\text{G}}$ and its time-reversal one are given by 
\begin{widetext}
\begin{gather}
H_{\text{G}} = v_f (p_x + e A_x(y)) \sigma_x + v_f (p_y + e A_y(y)) \sigma_y + s_j~ m(y), \nonumber\\
\mathcal{T} H_{\text{G}} \mathcal{T}^{-1} = 
v_f (p_x - e A_x(y)) \sigma_x + v_f (p_y - e A_y(y)) \sigma_y - s_j~ m(y), \label{GraH} 
\end{gather}
\end{widetext}
where the sign $s_{j=1,2}=1$ and $s_{j=3,4} = -1$ are for that of Zeeman energy.  
The position dependent superconducting gap $\Delta(y)$, vector potential $\vec{A}(y)=A_x(y) \hat{x}+
A_y(y) \hat{y}$, and Zeeman term $m(y)$ are written as
\begin{equation}\label{PositionDependence}
(\Delta(y), \vec{A}(y), m(y))= 
\left\{ 
 \begin{array}{cc}
  (\Delta_0 e^{i \phi}, 0, 0) \,& \hbox{$y < 0$},\\
  (0, -eBy \, \hat{x}, g \mu_B B/2) \,&  \hbox{$y > 0$},\\
 \end{array}
\right.
\end{equation}
where $g$ is the $g$-factor, $\mu_B$ Bohr magneton, $B>0$ is the strength of the magnetic field perpendicular to the graphene layer, and $- e$ is electron charge. We note that this graphene-superconductor interface was studied in Ref.~\cite{Akhmerov_graphene} for the case of zero Zeeman energy. 

The wave function in the superconducting region is 
\begin{widetext}
\begin{eqnarray}
\Psi_j (y<0) &=& 
a_{1j} ~ e^{i k_x x} e^{i k_{y1} y}
\left(
 \begin{array}{c}
  e^{-i \phi/2} (-\Delta_0) \hbar v_f (i k_x + k_{y1})\\
  e^{-i \phi/2} \Delta_0 (\mu + i \sqrt{\Delta^{2}_0 - \epsilon_j^2}) \\
  e^{i \phi/2} \hbar v_f (i k_x + k_{y1}) (-\epsilon_i + i \sqrt{\Delta^{2}_0 - \epsilon_j^2} )\\
  e^{i \phi/2} (\mu + i \sqrt{\Delta^{2}_0 - \epsilon_j^2} ) (\epsilon_j - i \sqrt{\Delta^{2}_0 - \epsilon_j^2} )
 \end{array}
\right) \nonumber\\
&+& a_{2j} ~ e^{i k_x x} e^{i k_{y2} y}
\left(
 \begin{array}{c}
  e^{-i \phi/2} (-\Delta_0) \hbar v_f (i k_x + k_{y2})\\
  e^{-i \phi/2} \Delta_0 (\mu - i \sqrt{\Delta^{2}_0 - \epsilon_j^2}) \\
  e^{i \phi/2} \hbar v_f (i k_x + k_{y2}) (-\epsilon_j - i \sqrt{\Delta^{2}_0 - \epsilon_j^2} )\\
  e^{i \phi/2} (\mu - i \sqrt{\Delta^{2}_0 - \epsilon_j^2} ) (\epsilon_j + i \sqrt{\Delta^{2}_0 - \epsilon_j^2} )
 \end{array}
\right),\label{WaveftnS}
\end{eqnarray}
\end{widetext}
where $k_x$ is the wave vector along the interface, $v \hbar k_{y 1} = - \text{sgn}(\mu) \sqrt{\left ( \mu +  i \sqrt{\Delta^{2}_0-\epsilon_j^2} \right)^2 - \hbar^2 v_f^2 k^{2}_x}$, $v \hbar k_{y 2} = \text{sgn}(\mu) \sqrt{\left ( \mu -  i \sqrt{\Delta^{2}_0-\epsilon_j^2} \right)^2 - \hbar^2 v_f^2 k^{2}_x}$, and $a_{1j}$ and $a_{2j}$ are coefficients.
In the graphene region, the electron wave function $\Psi^{e}_j(y>0)$ and hole $\Psi^{h}_j(y>0)$ are 
\begin{widetext}
\begin{gather}
\Psi^{e}_j(y>0) =
b^{e}_j ~ e^{i k_x x} e^{-(y - l^{2}_B k_x)^2/(2 l^{2}_B)}
\left(
 \begin{array}{c}
  \frac{i(\epsilon_j +\mu - s_j g \mu_B B/2)}{\sqrt{|\hbar v_f^2 e B|}} \text{H}_{u_e -1}(y/ l_B - l_B k_x)  \\
  \text{H}_{u_e}(y/ l_B - l_B k_x) \\
 \end{array}
\right), \nonumber\\ 
\Psi^{h}_i(y>0) =  
b^{h}_j ~ e^{i k_x x} e^{-(y + l^{2}_B k_x)^2/(2 l^{2}_B)}
\left(
 \begin{array}{c}
   \text{H}_{u_h}(y/ l_B + l_B k_x) \\
  \frac{-i(\epsilon_j -\mu -s_j g \mu_B B/2)}{\sqrt{|\hbar v_f^2 e B|}} \text{H}_{u_h -1}(y/ l_B + l_B k_x)
 \end{array}
\right),\label{WaveftnTI}
\end{gather}
\end{widetext}
where $u_{e(h)} = (\epsilon_j -s_j m  \pm \mu)^2/|2\hbar v_f^2 e B|$, the sign $+(-)$ for electron (hole) state, and $b_j^e$ and $b_j^h$ are coefficients.

\subsection*{2. Low-energy dispersion}

In this section, from the boundary matching of the wave functions in Eqs.~\eqref{WaveftnS} and \eqref{WaveftnTI} at $y=0$, we derive energy dispersion of chiral edge states near the zero excitation energy in the graphene case. In the low-energy regime of $|\epsilon_j| \ll |\mu|,|\Delta_0|$, the chiral edge states have the linear dispersion of the form 
\begin{equation}
\epsilon_j(k_x) = \hbar v_D k_x + E_{Z,j} + \mathcal{O}(k_x^2), \label{E_dispersion}
\end{equation}
where $v_D$ is the propagation velocity, and $E_{Z,j}$ is the contribution of Zeeman energy.
Due to the sign of the Zeeman energy, shown in Eq.~\eqref{GraH}, 
$\epsilon_1 (k_x) = \epsilon_2 (k_x)$ and $\epsilon_3 (k_x) = \epsilon_4 (k_x)$ 
while $\epsilon_{j=1,2} (k_x) \ne \epsilon_{j=3,4} (k_x)$. 
The Zeeman-energy dependence of $v_D$ is negligible in our parameters of $|g \mu_B B/2| = 0.06 \text{meV} \ll \mu = 15 \text{meV}$ at $B = 1 \,\, \text{T}$ and for at $g = 2$; the weak dependence does not change our finding in the main text that the interferometry setup based on graphene shows the $\Phi$-dependence in the zero bias limit. 

The detailed expression of the Zeeman contribution $E_{Z,j}$ is obtained from the equation for $\epsilon_{j,0} \equiv \epsilon_j (k_x = 0)$ [see Eq.~\eqref{E_dispersion}], which is derived from the boundary matching of $\Psi_j (y)$ at $y=0$. The equation for $\epsilon_{j,0}$ is 
\begin{widetext}
\begin{equation}
\epsilon_{j,0} + \frac{\sqrt{\Delta_0^2 - \epsilon_{j,0}^2}}{\sqrt{|\hbar v_f^2 e B|}} \left(
\left( \epsilon_{j,0} + \mu - \frac{s_j g \mu_B B}{2} \right) f_e + \left( \epsilon_{j,0} - \mu - \frac{s_j g \mu_B B}{2} \right) f_h \right) 
- \frac{(\epsilon_{j,0} - \frac{s_j g \mu_B B}{2})^2 - \mu^2}{|\hbar v_f^2 e B|} \epsilon_{j,0} f_e f_h = 0, 
\end{equation}
\end{widetext}
where $f_{e(h)} = \text{H}_{u_{e(h)}-1}(0)/\text{H}_{u_{e(h)}}(0)$ and
$\text{H}_{u_{e(h)}}(0)$ is the Hermite function with $u_{e(h)} = (\epsilon_j - s_j g \mu_B B / 2 \pm \mu )^2/|2 \hbar v^2_f e B|$. 
By taking into account of the low-energy limit of $|\epsilon_{j,0}/\mu|, |\epsilon_{j,0}/\Delta_0| \ll 1$, and by replacing $\epsilon_{j,0}$ [$\equiv \epsilon_j (k_x = 0)$] to $E_{Z,j}$, this equation is simplified to be 
\begin{widetext}
\begin{equation}
E_{Z,j} + \frac{\Delta_0}{\sqrt{|\hbar v_f^2 e B|}} \left(
\left( E_{Z,j} + \mu - \frac{s_j g \mu_B B}{2} \right) f^0_e + \left(E_{Z,j} - \mu - \frac{s_j g \mu_B B}{2} \right) f^0_h \right) 
- \frac{(\frac{g \mu_B B}{2})^2 - \mu^2}{|\hbar v_f^2 e B|} E_{Z,j} f^0_e f^0_h = 0, 
\end{equation}
\end{widetext}
where $f^0_{e(h)} = \text{H}_{u^0_{e(h)}-1}(0)/\text{H}_{u^0_{e(h)}}(0)$ and $u^0_{e(h)} = (\mu \mp s_j g \mu_B B / 2)^2/|2 \hbar v^2_f e B|$. This equation is rewritten as
\begin{equation}
E_{Z,i} = \frac{-(\mu -s_i g \mu_B B / 2) f_e +(\mu + s_i g \mu_B B / 2) f_h}{f_e + f_h + \frac{\sqrt{|\hbar v^2_f e B|}}{\Delta_0} +\frac{[\mu^2 - (g \mu_B B / 2)^2] f_e f_h}{\Delta_0 \sqrt{|\hbar v^2_f e B|}}}.
\end{equation}
Note that $|E_{Z,j}| = 6 \,\, \mu \text{eV}$ at $B = 1 \,\, \text{T}$ under the parameters (such as $\mu$) used in the main text.

%

\subsection*{3. Differential conductance of the interferometry setup based on graphene}

In this section, we provide the form of differential conductance $dI/dV$ of the interferometry based on graphene.
At zero temperature, the differential conductance $dI/dV$ from the lead to the four chiral edge channels has the form (which is a trivial generalization of Eq. (6) of the main text into the case of four channels) of
\begin{equation}
\frac{dI}{dV} = \sum^{4}_{j=1} \frac{2e^2}{h} |s^j_{he}|^2 = \sum^{4}_{j=1} \frac{2e^2}{h} \frac{\tilde{t}^4_l \text{cos}^2(\theta^j_s/2)}{\text{sin}^2(\theta^j_s/2)+\tilde{t}^4_l \text{cos}^2(\theta^j_s/2)},
\end{equation}
where $j$ represents the index of the chiral edge channels. 
Here, we assume that the chiral edge channels do not mix each other in the interferometry setup and at the Josephson junction, and 
each channel couples to the normal lead with the tunneling strength $t_l$ of $\tilde{t}_l = t_l/(2 \hbar \sqrt{v_D v_L})$. In this case, the Andreev reflection amplitudes satisfy $s^1_{he} = s^2_{he}$ and $s^3_{he} = s^4_{he}$, because of 
$\epsilon_1 = \epsilon_2$ and $\epsilon_3 = \epsilon_4 $ which are mentioned above. 
And, in the same way with the main text, we obtain $\theta^j_s = k_j L + \pi + n_v \pi + \theta^j_{\Lambda}$ (which is a trivial generalization of Eq. (5) of the main text), where $k_j = (\text{eV} - E_{Z,j})/(\hbar v_D)$ and
the phase shift $\theta^j_{\Lambda}$ by the Josephson junction is given by
\begin{equation}
\theta^j_{\Lambda} = \frac{r + (r^2 + t^2) e^{i k_j d}}{1 + r e^{i k_j d}}.
\end{equation}
The coefficients of reflection $r$ and transmission $t$ are the same as those of Eq.(3) in the main text. Here, we neglect the change of energy dependence of $r$ and $t$ by the Zeeman contribution $E_{Z,j}$, since the change is negligible in our parameter regime where $|E_{Z,j}/\Delta_0| \sim 4 \times 10^{-3}$.


\begin{thebibliography}{99}

\bibitem{Kitaev} A. Kitaev, Phys. Usp. \textbf{44}, 131 (2001).

\bibitem{Hasan} M. Z. Hasan and C. L. Kane, Rev. Mod. Phys. \textbf{82}, 3045 (2010).

\bibitem{Alicea} J. Alicea, Rep. Prog. Phys. \textbf{75}, 076501 (2012).

\bibitem{Flensberg} M. Leijnse and K. Flensberg, Semicond. Sci. Technol. {\bf 27}, 124003 (2012).

\bibitem{Mourik} V. Mourik, K. Zuo, S. M. Frolov, S. R. Plissard, E. P. A. M. Bakkers, and L. P. Kouwenhoven, Science {\bf 336}, 1003 (2012).

\bibitem{Das} A. Das, Y. Ronen, Y. Most, Y. Oreg, M. Heiblum, and H. Shtrikman, Nat. Phys. {\bf 8}, 887 (2012).

\bibitem{Deng} M. T. Deng, C. L. Yu, G. Y. Huang, M. Larsson, P. Caroff, and H. Q. Xu, Nano Lett. {\bf 12}, 6414 (2012).


\bibitem{Rokhinson} L. P. Rokhinson, X. Liu, and J. K. Furdyna, Nat. Phys. {\bf 8}, 795 (2012).

\bibitem{Oreg} Y. Oreg, G. Refael, and F. von Oppen, Phys. Rev. Lett. {\bf 105}, 177002 (2010).

\bibitem{Lutchyn} R. M. Lutchyn, J. D. Sau, and S. Das Sarma, Phys. Rev. Lett. {\bf 105}, 077001 (2010).

\bibitem{Williams} J. R. Williams, A. J. Bestwick, P. Gallagher, S. S. Hong, Y. Cui, A. S. Bleich, J. G. Analytis, I. R. Fisher, and D. Goldhaber-Gordon, Phys. Rev. Lett \textbf{109}, 056803 (2012).

\bibitem{Sacepe} B. Sac\'{e}p\'{e}, J. B. Oostinga, J. Li, A. Ubaldini, N. J. G. Couto, E. Giannini and A. F. Morpurgo, Nature Comm. \textbf{2}, 575, (2011).

\bibitem{Veldhorst} M. Veldhorst, M. Snelder, M. Hoek, T. Gang, V. K. Guduru, X. L. Wang, U. Zeitler, W. G. van der Wiel, A. A. Golubov, H. Hilgenkamp and A. Brinkman, Nature Mat. \textbf{11}, 417 (2012).

\bibitem{Fu} L. Fu and C. L. Kane, Phys. Rev. Lett. \textbf{100}, 096407 (2008).

\bibitem{Fu_interf} L. Fu and C. L. Kane, Phys. Rev. Lett. \textbf{102}, 216403 (2009).

\bibitem{Akhmerov_interf} A. R. Akhmerov, J. Nilsson, and C. W. J. Beenakker, Phys. Rev. Lett. \textbf{102}, 216404 (2009).

\bibitem{Law} K. T. Law, P. A. Lee, and T. K. Ng, Phys. Rev. Lett. \textbf{103}, 237001 (2009).

\bibitem{Diez} M. Diez, I. C. Fulga, D. I. Pikulin, M. Wimmer, A. R. Akhmerov, and C. W. J. Beenakker, Phys. Rev. B \textbf{87}, 125406 (2013).

\bibitem{Wieder} B. J. Wieder, F. Zhang, and C. L. Kane, arXiv:1302.2113.

\bibitem{Beri} B. Beri, Phys. Rev. B {\bf 85}, 140501(R) (2012).

\bibitem{Tanaka} Y. Tanaka, T. Yokoyama, and N. Nagaosa, Phys. Rev. Lett. {\bf 103}, 107002 (2009).

\bibitem{Li} J. Li, G. Fleury, and M. B\"{u}ttiker, Phys. Rev. B {\bf 85}, 125440 (2012).

\bibitem{Greenberger} D. M. Greenberger, D. K. Atwood, J. Arthur, C. G. Shull, and M. Schlenker, 
Phys. Rev. Lett. \textbf{47}, 751 (1981).

\bibitem{Fu_tele} L. Fu, Phys. Rev. Lett. {\bf 104}, 056402 (2010).

\bibitem{Ji} Y. Ji, Y. Chung, D. Sprinzak, M. Heiblum, D. Mahalu, and H. Shtrikman, Nature {\bf 422}, 415 (2003).

\bibitem{Akhmerov_graphene} A. R. Akhmerov and C. W. J. Beenakker, Phys. Rev. Lett. \textbf{98}, 157003 (2007).

\bibitem{Tiwari} R. P. Tiwari, U. Z\"{u}licke, and C. Bruder, Phys. Rev. Lett. \textbf{110}, 186805 (2013).

\bibitem{Park} S. Park and H.-S. Sim, in preparation.

\bibitem{Deviation_real} For $\hbar v_M k \ll \Delta_0$, $r \simeq r_0 [1 + i \frac{r_0 k W_\Delta}{\cos (2 \pi \Phi / \Phi_{0,e})}]$ and $r^2 + t^2 \simeq 1 + i \frac{ 2 r_0 k W_\Delta}{\cos (2 \pi \Phi / \Phi_{0,e})}$, where $W_\Delta \equiv \hbar v_M / \Delta_0$. 

\bibitem{Rakhmanov} A. L. Rakhmanov, A. V. Rozhkov, and Franco Nori, Phys. Rev. B \textbf{84}, 075141 (2011).
\bibitem{Ioselevich} P. A. Ioselevich, P. M. Ostrovsky, and M. V. Feigel'man, Phys. Rev. B \textbf{86}, 035441 (2012);
P. A. Ioselevich, and M V Feigel'man, New J. Phys. \textbf{15}, 055011 (2003).

\bibitem{Geim} A. H. Castro Neto, F. Guinea, N. M. R. Peres, K. S. Novoselov, and A. K. Geim, Rev. Mod. Phys. {\bf 81}, 109 (2009).


\bibitem{Zeeman} See Supplemental Material for the details.

\bibitem{Rickhaus} P. Rickhaus, M. Weiss, L. Marot, and C. Sch\"{o}nenberger, Nano Lett. \textbf{12}, 1942 (2012).

\bibitem{Morpurgo2} A. F. Morpurgo, J. Kong, C. M. Marcus, H. Dai, Science \textbf{286}, 263 (1999). 

\bibitem{Oostinga} J. B. Oostinga et. al., Phys. Rev. X \textbf{3}, 021007 (2013).

\bibitem{Eroms} J. Eroms, D. Weiss, J. De Boeck, G. Borghs, and U. Z\"{u}licke, Phys. Rev. Lett. \textbf{95}, 107001 (2005).

\bibitem{Uhlisch} D. Uhlisch, S. G. Lachenmann, Th. Sch\"{a}pers, A. I. Braginski, H. L\"{u}th, J. Appenzeller, A. A. Golubov, A. V. Ustinov, Phys. Rev. B \textbf{61}, 12463 (2000).

\end{thebibliography}
\end{document}